\documentclass[conference]{IEEEtran}
\IEEEoverridecommandlockouts
\usepackage{comment}
\usepackage{cite}
\usepackage{amsmath,amssymb,amsfonts}
\usepackage{algorithmic}
\usepackage{graphicx}
\usepackage{textcomp}
\usepackage{xcolor}
\usepackage{dblfloatfix} 
\usepackage{tabularx}
\usepackage{textcomp}
\usepackage{xcolor}
\usepackage{lipsum}
\usepackage{booktabs}
\usepackage{multirow}
\usepackage{algorithmic}
\usepackage{graphicx}
\usepackage{algorithm}
\usepackage{enumitem}
\usepackage{bm}

\usepackage{tikz}
\usetikzlibrary{positioning}
\usetikzlibrary{decorations.pathreplacing}
\usepackage{amsmath}
\def\BibTeX{{\rm B\kern-.05em{\sc i\kern-.025em b}\kern-.08em
    T\kern-.1667em\lower.7ex\hbox{E}\kern-.125emX}}
\begin{document}
%\listoftodos[A list of things I need to finish]
\title{EMGTFNet: Fuzzy Vision Transformer to decode Upperlimb sEMG signals for Hand Gestures Recognition
%-based Hand Gesture Recognition from High Density Surface EMG Signals
%Decoding of Upperlimb sEMG for Hand Gestures Recognition
%*\\
%5\thanks{Identify applicable funding agency here. If none, delete this.}
}

\author{\IEEEauthorblockN{1\textsuperscript{st} Joseph Cherre Córdova}
\IEEEauthorblockA{\textit{Dept. of Electrical Engineering} \\
\textit{Universidad de Ingenieria y Tecnologia}\\
Lima, Peru \\
joseph.cherre@utec.edu.pe }
\and

\IEEEauthorblockN{2\textsuperscript{nd} Christian Flores}
\IEEEauthorblockA{\textit{Dept. of Electrical Engineering} \\
\textit{Universidad de Ingenieria y Tecnologia}\\
Lima, Peru \\
\textit{University of Campinas}\\
São Paulo, Brazil\\
cflores@utec.edu.pe }
\and

\IEEEauthorblockN{3\textsuperscript{rd} Javier Andreu-Perez}
\IEEEauthorblockA{\textit{Centre for Computational Intelligence} \\
\textit{University of Essex}\\
Colchester, United Kingdom \\
j.andreu-perez@essex.ac.uk}

}

\maketitle

\begin{abstract}
 %\todo[color=blue!40]{Josep Básate en este paper: https://arxiv.org/abs/2201.10060 para escribir el artículo}
Myoelectric control is an area of electromyography of increasing interest nowadays, particularly in applications such as Hand Gesture Recognition (HGR) for bionic prostheses. The focus today is on pattern recognition using Machine Learning and, more recently, Deep Learning methods. Despite achieving good results on sparse sEMG signals, the latter models typically require large datasets and training times. Furthermore, due to the nature of stochastic sEMG signals, traditional models fail to generalize samples for atypical or noisy values. In this paper, we propose the design of a Vision Transformer (ViT) based architecture with a Fuzzy Neural Block (FNB) called EMGTFNet to perform Hand Gesture Recognition from surface electromyography (sEMG) signals.
The proposed EMGTFNet architecture can accurately classify a variety of hand gestures without any need for data augmentation techniques, transfer learning or a significant increase in the number of parameters in the network. The accuracy of the proposed model is tested using the publicly available NinaPro database consisting of 49 different hand gestures. Experiments yield an average test accuracy of 83.57\% ± 3.5\% using a 200 ms window size and only 56,793 trainable parameters. Our results outperform the ViT without FNB, thus demonstrating that including FNB improves its performance. Our proposal framework EMGTFNet reported the significant potential for its practical application for prosthetic control.

%The compact structure of the proposed ViT-based ViT-HGR framework (i.e., having significantly reduced number of trainable parameters) shows great potentials for its practical application for prosthetic control.

%Although these tests were performed offline, the computation should be fast enough to implement in real-time prosthetic control systems. Future work remains to implement this algorithm in an online prosthetic control system, as it shows potential to be a promising application in the field.

\end{abstract}


\section{Introduction}
 %\todo[color=green!40]{Christian}%
Control Prosthesis demands effective tasks classification
to perform according to the end-user requirements. Moreover, this system has to overcome the difficulties presented in real-life conditions such as fatigue, noise environment, etc\cite{Iqbal_2018}. Thus, they need to reach outstanding results to avoid unexpected actions that may cause rejection from the patients. Therefore, additional activity tasks are essential to give the patients more versatility during daily activities \cite{Ninapro_database_info_2}.

Mainly, the Control Prosthesis is performed by electric biosignals from muscles. Hence,  Electromyography (EMG) is a technique for assessing and recording the electrical activity produced by muscles. These Electric biosignals have been studied to detect muscle movements, mainly using non-invasive techniques due to their rapid and low-risk procedure \cite{Yang_2022}.
One essential step to classify the EMG signal is feature engineering, ergo obtaining characteristic features to improve classification accuracy. 
These features can be obtained from surface electromyographic (sEMG) signals, which are bioelectric impulses obtained through a non-invasive procedure. They enable the identification of movements, actions, and orientations \cite{Reaz_2006}. 

One approach to analyzing the sEMG is identifying patterns over time, such as fractal dimension analysis. For instance, in  \cite{Escandon_2021_Springer}, \cite{Escandon_IEEE} Higuchi's fractal dimension was applied in the pre-processing stage to classify sEMG signals for different hand movements. Another approach is identifying patterns in the time and frequency domain to take advantage of their frequency composition \cite{Atzori_2014}. %Some studies reported suitable classification accuracy estimations with several classification techniques \cite{Robinson_2017}. %Different techniques of machine learning and feature enengineering have reported interesting results. For instance, in \cite{Atzori_2014} the authors used Root-Mean Square (RMS), Discrete Wavelength Transform (DWT) in combination with linear Random Forest (RF) to analyze and classify EMG samples. Their results reported 75.32\% of classification accuracy for 50 Hand Gesture Recognition (HGR). 
%Similarly, in this paper \cite{Zhai_2016}, the author used a dataset of  40 subjects who performed 50 different hand movements. Therefore, EMG Spectrograms were calculated to turn into signals onto images. Also, the Principal Component Analysis (PCA) to reduce the dimension was performed. Finally, a Support Vector Machine (SVM) classifies the 50 hand movements achieving 76\% of classification accuracy.
Moreover, Deep Learning (DL) is an alternative method for hand gesture classification that can significantly improve accuracy in classification. 
Among the most prevalent approach is a Convolutional Neural Network (CNN) \cite{Rahimian_2019} due to the power to learn spatial features. However, CNN is not able to extract temporal features, which is the principal characteristic of EMG signals. To overcome this issue Recurrent Neural Networks (RNNs) and Long Short Term Memory (LSTM) have been proposed in recent studies \cite{Sun_2021} to extract temporal information from sEMG signals. Therefore, a hybrid solution can be proposed to take advantage of two kinds of networks (LSTM and CNN). 
One disadvantage of these sequence model approaches is that they do not allow parallelization; hence their train slow and difficult to train. To tackle these problems, a Transformer was proposed in \cite{Vaswani_2017}. A transformer is a deep learning model using a self-attention mechanism instead of recurrence or convolution. It differentially weights the significance of each part of the input data. %The Transformer neural network architecture depicts a powerful DNN model using an attention mechanism to select weighting the input information.
For instance, in \cite{Elahe_2021} proposed a Vision Transformer-based neural network called TEMGNet to classify and recognize upper limb hand gestures from sEMG. Using a NinaPro DB2 dataset (exercise B), their results reached a classification accuracy of 82.93\% and 82.05\% for window sizes of 300ms and 200ms. Nevertheless, EMG patterns exhibit high variability across sessions and uncertainty due to noise. A drawback of learning EMG signals with DL is the assumption that these patterns are noise-free and not affected by non-stationarity, ignoring uncertainty. This assumption is critical as we can control prostheses in real environments. In this regard, fuzzy sets and systems have reported suitable performance in noisy environments.
\cite{Vega_2022}

Motivated by the success of Transformers in classification problems of  EMG signals, we propose designing a Vision Transformer  (ViT) architecture with a fuzzy neural block (FNB) in parallel in the linear layer and the last layer before the output of the encoder. To aim our objective, we choose a TEMGNet \cite{Elahe_2021} to assess the inclusion of FNB. We tested our proposed deep learning model’s efficacy with FNB to classify the NinaPro DB2 dataset (exercise B).
%The paper addresses this gap and focuses on the design of a Fuzzy transformer.
The main contributions of the present work are:

\begin{itemize}
\item To the best of our knowledge, this is the first paper that develops a Fuzzy Transformer based architecture for the task of HGR using sEMG.

\item A proposal of a Fuzzy Transformer architecture was assessed for 17 HGR from 40 subjects. The results indicate higher accuracy of the proposed model versus state-of-the-art methods.
\end{itemize}
%Our results 
%The paper demonstrates the superior performance of the
%propose hybrid architecture TraHGR and its ability to extract
%more distinctive information for gesture recognition
%over the single models; i.e., TNet and FNet.

%More specifically, compared
%with the proposed architectures in the recent
%state-of-the-art studies, TraHGR improves the recognition
%accuracies to 86:18\% on DB2 (49 gestures),
%to 88:91\% on the DB2-B (17 gestures), to 81:44\%
%on DB2-C (23 gestures), and to 93:84\% on the
%DB2-D (9 gestures).
This work is organized as follows: Section II presents the materials and methods. The description and proposal for Fuzzy Transformer are shown in Section III. The description of the experiment and results are discussed in Section IV. Finally, in Section V, we present the conclusions.

\section{Materials and Methods}

\subsection{The Dataset}
\begin{figure}
    \centering
    \includegraphics[width=230px,height=370px,keepaspectratio]{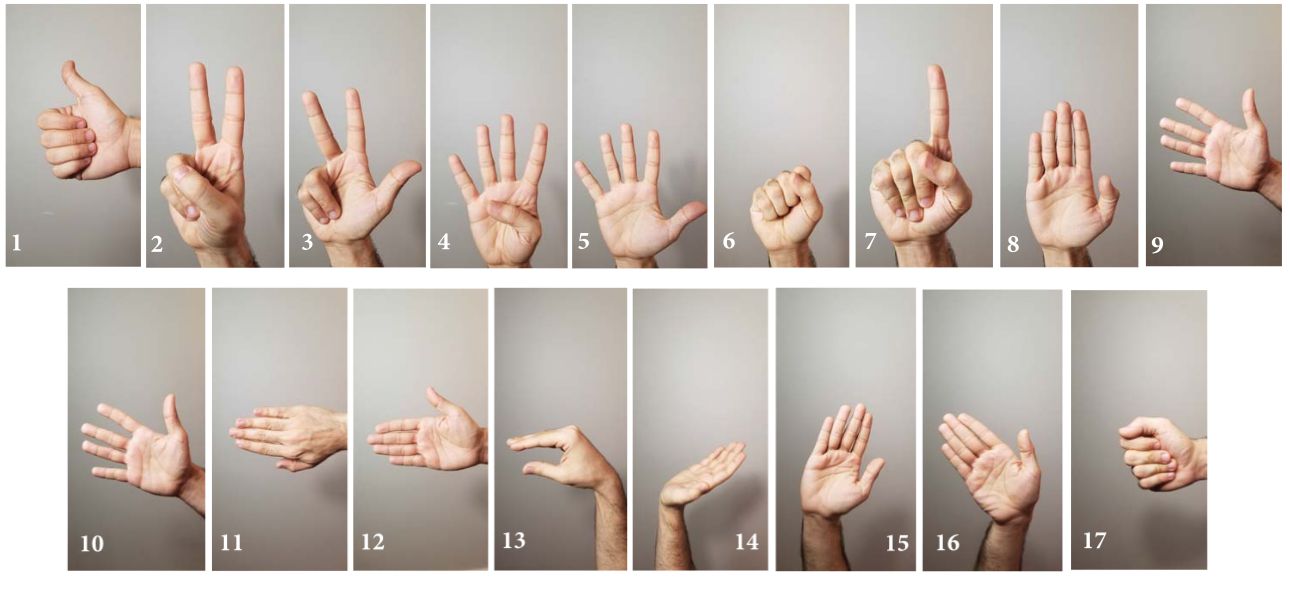}
    \caption{ Exercise B of Ninapro database \cite{Ninapro_database_info_2}}
    \label{fig:DB2-B}
\end{figure}

%\todo[color=blue!40]{Josehp hablar de la base de datos de Nina Pro.}
%\cite{Atzori_2014} 
%Ninapro_database_info_2
%NinaPro
The proposed model was evaluated on the second Ninapro benchmark database \cite{Ninapro_database_info_2},  referred to as the DB2. This publicly available dataset contains sparse multichannel sEMG data widely used to classify hand gestures \cite{Sun_2021}, \cite{Elahe_2021}. For its construction, surface EMG signals were collected from 40 healthy individuals (12 females and 28 males ranging in age from 29.9 ± 3.9) who repeated several movements 6 times, each lasting 5 seconds, followed by 3 seconds of rest. The electrical signals from muscular activities were measured through 12 Delsys Trigno wireless electrodes at 2kHz of sampling frequency. Users perform 49 different types of movements, such as hand gestures, functional gestures, wrist gestures and grasping movements combined with different force patterns. Gestures in the DB2 dataset are divided into three exercises (i.e., B, C, and D). For comparison purposes, this paper evaluates the proposed model using Exercise B, which consists of 17 different movements, 9 basic wrist movements with 8 isometric and isotonic hand configurations, as shown in Fig. \ref{fig:DB2-B}. In order to compare and follow the recommendations provided by the Ninapro database, 2/3 of the gesture repetitions of each user (i.e., 1, 3, 4, and 6) were used to build the training set, and the remaining repetitions (i.e., 2, and 5) were used for testing.

\subsection{Initial Preprocessing}
In the present study, the open-source web interface Jupyter based on the Python language is used to develop the processing and classification algorithms using the Pytorch package.
The following steps are performed as data preparation to use the proposed DL architecture. The first step consists of removing power line interference. This filtering is performed using a notch filter or a Hampel filter. Based on other authors, on the one hand, it is suggested to avoid the notch filter because the power line frequency overlaps with the region of the spectrum where the peak of the sEMG signal is located. On the other hand, the Hampel filter calculates the median and standard deviation in a sliding window and replaces the outliers with the median value. Then, a Hampel filter is used in the 50 Hz band. Subsequently, to reduce interference from motion artifacts and high-frequency noise, multiple digital filters are applied, namely, a 20 Hz high-pass filter and a 500 Hz low-pass filter. The random nature of the EMG signal means that it cannot be reproduced exactly a second time. Therefore, the raw signal is not useful for pattern recognition algorithms. Instead, a smoothing filter is applied to extract a signal that is an estimator of the amplitude behavior. For this, an RMS filter is used according to the following equation:

\begin{equation}
\label{RMSFilter}
RMS_{k} = \sqrt{\frac{1}{k}\sum_{i=1}^{k}x_{i}^2}
\end{equation}

Similar to the mean absolute value (MAV) filter, a moving average is calculated based on calculating the root mean square of the mean square of the \(k\) sEMG samples corresponding to a window. The window size is usually selected between 50ms and 250ms to avoid phase shifts in abrupt changes of the signal \cite{tsinganos_2021}. The window size used in the present work for RMS calculation is 100ms. After the extraction of the RMS envelope, the signals are undersampled at 100 Hz and then a low-pass filter with a cutoff frequency of 1 Hz is applied, as suggested in \cite{Atzori2016}.

Subsequently, a segmentation scheme using overlapping rectangular windows with a window size of 200 ms and an overlap of 10 ms between successive windows is adopted. In this scenario, the EMG-PR system requires (150 + $\tau$) ms to predict the desired gesture, where $\tau$ is the processing time of a classifier. Therefore the processing time required for EMG-PR is less than the recommended limit \cite{delay}.

Finally, the post-filtered signals pass through a normalization step presented in \cite{FSHGR}. This normalization technique is defined as \(\mu\)-law normalization and is shown in the following equation:

\begin{equation}
\label{ulaw_Norm}
F(x_{t}) = sign(x_{t})\frac{\ln{1+\mu|x_{t}|}}{\ln{1+\mu}}
\end{equation}

Where \(x_{t}\) represents the input scalar, and the parameter \(\mu\) denotes the new rank. Recently, in \cite{Elahe_2021}, it was shown that the performance could be improved using normalization of the sEMG signals with the \(\mu\)-law approach.

\section{The proposed EMGTFNet Framework}
%\todo[color=blue!40]{Josehp hablar de la propuesta.}
The proposed EMGTFNet framework's details are described in this section for further explanation. Of particular interest for this work is the architecture known as Vision Transformer \cite{vision_transformer}, derived from the original Transformer used for machine translation and natural language processing (NLP) tasks in \cite{transformer}. The architecture is primarily based on the attention mechanism and does not take into account recurrence and convolutions. This mechanism allows the network to establish global dependencies between input and output. The standard Transformer architecture consists of an encoder block and a decoder block; however, the Vision Transformer architecture is based on the encoder block only. The proposed framework's main contribution is adding a Fuzzy Neural Block to the ViT based architecture used in \cite{Elahe_2021} for hand gesture recognition. The two main modules of the frameworks are described in the following subsections.

\subsection{TEMGNet}
\begin{figure*}
    \centering
    \includegraphics[width=\textwidth,height=400px,keepaspectratio]{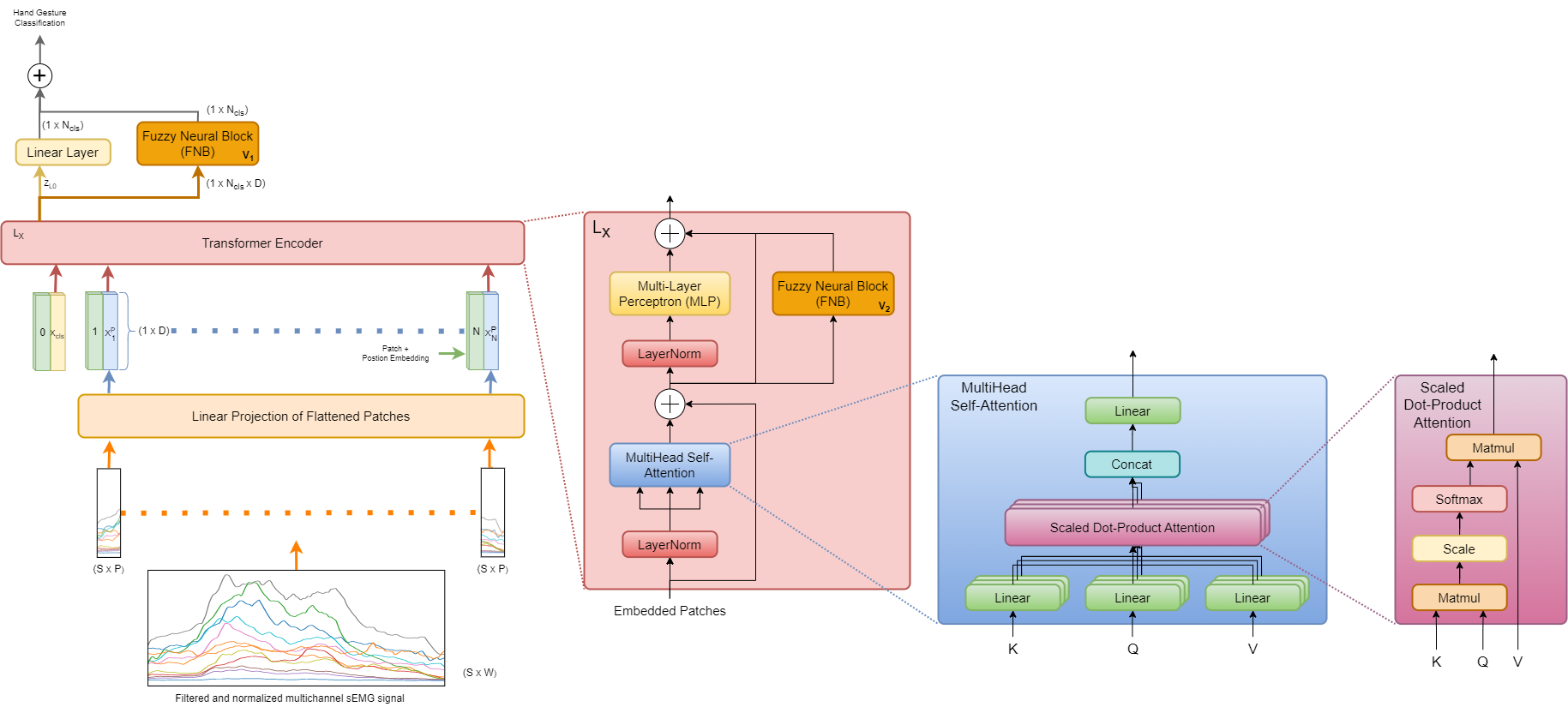}
    \caption{EMGTFNet: Proposed architecture based on the Vision Transformer (ViT) including a Fuzzy Neural block (FNB)}
    \label{figure:architecture_EMGTFNet}
\end{figure*}

The TEMGNet architecture was initially proposed in \cite{Elahe_2021}, and it showed promising results compared to the traditional convolution-based deep learning approaches. Attention mechanism is the core block of the Transformer architecture due to its proven ability to learn deep features and leverage the temporal information of the sEMG signals \cite{Hu2018}.
%Recently, attention mechanism has been proposed along with other architectures such as CNN or LSTM to classify hand gestures using sEMG signals. Nonetheless, the aim of this paper is to show that the proposed architecture based on the Vision Transformer (ViT), has the ability to outperform previous networks without the need for any data augmentation or feature extraction.
The general structure of the proposed architecture is shown in Fig. \ref{figure:architecture_EMGTFNet}

The main difference between the TEMGNet architecture and the original Vision Transformer used in \cite{vision_transformer} for image recognition tasks is in the initial segmentation used. While in the original paper, the entire image is fed to the network and then divided into square patches, the proposed architecture uses normalized data segments as input and rectangular patches are applied to each sEMG input sequence. The overall structure consists of the following blocks: Patch Embedding, Position Embedding, Transformer encoder, and finally a Multi-Layer Perceptron (MLP) head.

The ViT is designed to have a 3D image as an input, which is then reshaped to a sequence of flat 2D patches with a predetermined length. In order to convert the sEMG signals into an acceptable input, an additional dimension is added to simulate the channels of an image. The final dimensions of the  The first stage of the architecture consist in dividing the input sequence into multiple sections utilizing a sliding window of 20 samples (200 ms considering the 100 Hz undersampled frequency) with a skip step equal to 1 sample (10 ms). After the windowing step, each of the sequences \(\mathbf{X}\) are divided into a number \(N\) of small patches or non-overlapping segments of size \((S \times P)\); where \(S\) equals the number of channels and \(P\) is the patch size. Therefore, the number of patches will be equal to \(N = W/P\) considering an input \(\mathbf{X} \in \mathbb{R}^{S\times W}\) where \(W\) refers to the number of samples in the respective window.
Then, each patch is flattened into a vector \(\mathbf{x}_{j}^{p} \in \mathbb{R}^{S^{2}}\) for \((1 \leq j \leq N)\). A linear projection is then applied to embed each vector in the \(d\) dimension of the model. For the linear projection, we use a matrix \(\mathbf{E} \in \mathbb{R}^{S^{2}\times d}\) that is shared between different patches. The result of this projection is called patch embeddingm in Eq. \ref{patch_emb}. The sequence of embedded patches is attached with a trainable token \(\mathbf{x}^{cls}\), to capture the corresponding class of the input sequence. Finally, we add position embeddings denoted by \(\mathbf{E}_{pos} \in \mathbb{R}^{(N+1) \times d}\), which will allow the transformer to capture positional information. The formula for the input of the transformer encoder \(\mathbf{Z}_{0}\) is computed as follows:

\begin{equation}
\label{patch_emb}
\mathbf{Z_{0}} = \left[{\mathbf{x}^{cls};\mathbf{x}_{1}^{p}\mathbf{E};\mathbf{x}_{2}^{p}\mathbf{E};...;\mathbf{x}_{N}^{p}\mathbf{E};}\right] + \mathbf{E}_{pos}
\end{equation}

The encoder block is similar to the main transformer encoder block proposed by \cite{transformer}. As shown in Fig. \ref{figure:architecture_EMGTFNet}, the transformer encoder consists of \(L\) identical layers. Each layer consists of two modules: a Multihead Self-Attention (MSA) mechanism and an MLP module. The MSA is based on the self-attention mechanism (SA), which aims to capture the interaction between different vectors in a given input sequence. Let us define the input sequence as \(\mathbf{Z} \in \mathbb{R}^{N \times d}\), which consists of \(N\) vectors, each with an embedding dimension \(d\). In this sense, the SA mechanism will try to find the interactions between the \(N\) vectors in \(\mathbf{Z}\) by means of three different matrices, namely, \(\mathbf{Q}\), \(\mathbf{K}\) and \(\mathbf{V}\) representing query, key and value respectively. The formula is as follows:

\begin{equation}
\label{eq:selfatt_out}
SA(\mathbf{Z}) = softmax(\frac{\mathbf{Q}\mathbf{K}^{T}}{\sqrt{dh}})\mathbf{V}
\end{equation}

For the MSA mechanism, the SA mechanism is applied \(h\) times in parallel, allowing the model to attend to different parts of the input sequence for each head independently. The final values are achieved as follows:

\begin{equation}
\label{eq:multiheadselfatt}
MSA(\mathbf{Z}) = \left[SA_{1}(\mathbf{Z});SA_{2}(\mathbf{Z});...;SA_{h}(\mathbf{Z}) \right]W^{MSA}
\end{equation}

The output from Eq. \ref{eq:multiheadselfatt} passes on to the MLP module consisting of two linear layers and a Gaussian error linear unit (GELU) activation function. To address the problem of degradation, a layer normalization \cite{layer_norm} is applied. Residual jump connections are placed between each layer. The transformer encoder equation is then as follows:

\begin{equation}
\label{encoder1}
\mathbf{Z_{l}^{'}} = MSA(LayerNorm(\mathbf{Z}_{l-1}))+\mathbf{Z}_{l-1}
\end{equation}

\begin{equation}
\label{encoder2}
\mathbf{Z_{l}} = MLP(LayerNorm(\mathbf{Z}_{l}^{'}))+\mathbf{Z}_{l-1}^{'}
\end{equation}

For \(l = 1,...,L\), the encoder output equation is denoted as follows:

\begin{equation}
\label{encoder3}
\mathbf{Z_{L}} = \left[{\mathbf{z}_{L0};\mathbf{z}_{L1};...;\mathbf{z}_{LN}}\right]
\end{equation}

Where \(\mathbf{z}_{L0}\) is used for classification. Finally, \(\mathbf{z}_{L0}\) is passed to a linear layer (LL) as shown in the Eq. \ref{eq:mlp_head}

\begin{equation}
\label{eq:mlp_head}
%\mathbf{y} = LL(LayerNorm(\mathbf{z_{L0}}))
\mathbf{y} = LL(LayerNorm(\mathbf{z}_{L0}))
\end{equation}

\subsection{Fuzzy neural block}
%-----------------------------------------------------------
%-----------------------------------------------------------
% In each of the listed deep learning architectures in section \ref{sec:BG} for P300 based BCI we stack a FNB to be learn by the deep neural network.
%In this section, we provide a formal definition of our proposed FNB.
%\textbf{
%\begin{definition}[Fuzzy neural block]
A fuzzy neural block (FNB) is defined as a sequence of processing layers that activate the antecedents of a fuzzy rule. It first takes the normalised output of the previous layer $\bf {\bm{O}}$ that is subsequently flattened as $\bm{v}_{{i}}=Vec(\bm{O}_{{i}})$ where 
$\bm{v}_{{i}}=[o_{1,1},...,o_{m,1},o_{1,2},...,o_{m,2},...,o_{1,n},...,o_{m,n}]^T$

%$\bm{v_{i}} = Vec(\bm{O_{i}})=[o_{1,1},...,o_{m,1},o_{1,2},...,o_{m,2},...,o_{1,n},...,o_{m,n}]^T$. 

%$\bf {\bm{v_{i}}}$ = $Vec(\bm{\bf {O_{i}}})$=$[o_{1,1},...,o_{m,1},o_{1,2},...,o_{m,2},...,o_{1,n},...,o_{m,n}]^T$. 

\begin{comment}
{\scriptsize
\begin{equation}
\label{eq:vec_}
%\textbf {\textit {v_{i}=Vec(O_{i})=[o_{1,1},...,o_{m,1},o_{1,2},...,o_{m,2},...,o_{1,n},...,o_{m,n}]^T.
%\emph \textbf v_{i}
%v_{\mathbf{\textit{i}}}
%{\mathbf{\textit{v}}}
\bm{v}_{{i}}=Vec(\bm{O}_{{i}})=[o_{1,1},...,o_{m,1},o_{1,2},...,o_{m,2},...,o_{1,n},...,o_{m,n}]^T
%{\textit{v}}
\end{equation}
}
\end{comment}

Next, the fuzzy clustering approach in Kilic et al. \cite{kilicc2007} is used to find a set of $K$ centroids of shape $\bm{c}^{k}$=$[c^{k}_1,c^{k}_2,...,c^{k}_d]$, using a collection $\bm{D}$ of $\bm{v}_{{i}}$ of layer outputs collected from the previous epoch. For the first epoch, the centroids are set to zero. The Gaussian membership value of $\bm{v}_{\bm{i}}$ is computed as 
\begin{align}
%\begin{equation} 
    \bm{\mu}_{i}^{k}    (\bm{v}_{i},\bm{c}^{k},\bm{a})=exp\left(-1/4((\bm{v}_{i}-\bm{c}^{k})^{2}/\bm{a}^{2}\right)
\end{align}
%\end{equation}
the scaling vector $\bm{a}$ is a parameter that is set to learn by the network, and the rule activation consists of a t-norm operator and normalisation step such:
\begin{align} \label{eq:output_FNB}
    o^k(\bm{v}_{i}) = \prod_{j=1}^{d}\mu_{i,j}^{k} \hspace{10mm} and \hspace{10mm} \widetilde{o}^{k}=o^{k}/{\textstyle \sum_{f=1}^{K}o^{f}}
\end{align} 
where $d$ is the dimension of the $\bm{\mu}_{i}^{k}$ vector, and $\bm{O'} = [\widetilde{o}^{1},...,\widetilde{o}^{K}]$ is the output of the FNB that is forwarded to next layer. Note the output dimension of the FNB is reduced to $K$.
%\end{definition}

\subsection{Fuzzy Transformer}
We proposed 3 variants of EMGTFNet and one ViT as the baseline for comparison purposes. The proposed Fuzzy structure is based on \cite{Vega_2022} since this structure enhanced the classification accuracy. We list the description of each transformer architecture below.
\begin{itemize}
\item TEMGNet: The base model is a Vision Transformer (ViT)-based neural network architecture.
\item  EMGTF-Net v1: Vision Transformer with fuzzy block parallel to the layer linear output (marked as \(V_{1}\) in Fig. \ref{figure:architecture_EMGTFNet}).
\item  EMGTF-Net v2: Vision Transformer with fuzzy block parallel to multilayer perceptron of the encoder (marked as \(V_{2}\) in Fig. \ref{figure:architecture_EMGTFNet}).
\item  EMGTF-Net v3: Vision Transformer with fuzzy block parallel to layer output linear and multilayer perceptron of the encoder (\(V_{1}\) and \(V_{2}\) in Fig. \ref{figure:architecture_EMGTFNet}).
\end{itemize}
 %The EMGTF-Net v3 encompasses the EMGTF-Net v1 and EMGTF-Net v2 as we can see in the Fig.\ref{figure:architecture_EMGTFNet}
%---------------------------------
\section{Experiment and results}
We assess our proposed Fuzzy transformer on all 17 hand gestures  for each one of 40 subjects from the second Ninapro
benchmark database, as known as the DB2. As described in the previous section, we presented three models considering different positions of FNB. 
%After EMG was classified, it was preprocessed to enhance its classification. 
%\begin{enumerate}
%    \item It applies a moving-average RMS of 100 ms \cite{tsinganos_2021}.
%    \item Signals are downsampled from 2000 Hz to 100 Hz to reduce computational cost \cite{Atzori2016}.
%   \item It applies a first-order Butterworth low filter with cut-off frequency of 1 Hz \cite{Geng2016}. 
%   \item It applies a $\mu$-law normalization over the data \cite{Elahe_2021}. 
%   \item It applies a wDindowing of 200 ms and overlapping of 10 ms over the data \cite{Elahe_2021}. 
%\end{enumerate}

For each model, the same preprocessing steps were applied. A window size of 20 samples (200 ms) with a skip step equal to 1 sample (10 ms) was chosen for testing due to comparison purposes. The model's hyperparameters such as patch size, depth, dimension, MLP size, and the number of heads were configured in all models as follows: (12, 4), 1, 64, 256, and 8, respectively. The patch size was chosen according to \cite{Shen2022} who did a preliminary study on the effect of the patch size on a similar architecture. The rest of the model's parameters are picked based on the original study in \cite{Elahe_2021}.

The optimal training parameters of the Transformer were found by preliminary analysis, observing the behaviour of the network when trained with various settings. In particular the optimal learning rate, number of epochs and scheduling were chosen by preliminary analysis. The optimal settings, used for all training in the final stage, are as follows:

\begin{itemize}
  \item Random initialization of weights and biases.
  \item Loss function: cross-entropy loss, implemented by PyTorch command criterion, which stacks the log-softmax algorithm and negative log-likelihood loss.
  \item Optimization algorithm: Adam with \(lr=0.0001\), \(b1=0.9\), \(b2=0.999\), \(weight\_decay=0.001\) and batch size of 512; batches are randomly rescaled at each epoch.
\end{itemize}

\begin{figure}
    \centering
    \includegraphics[width=230px,height=370px,keepaspectratio]{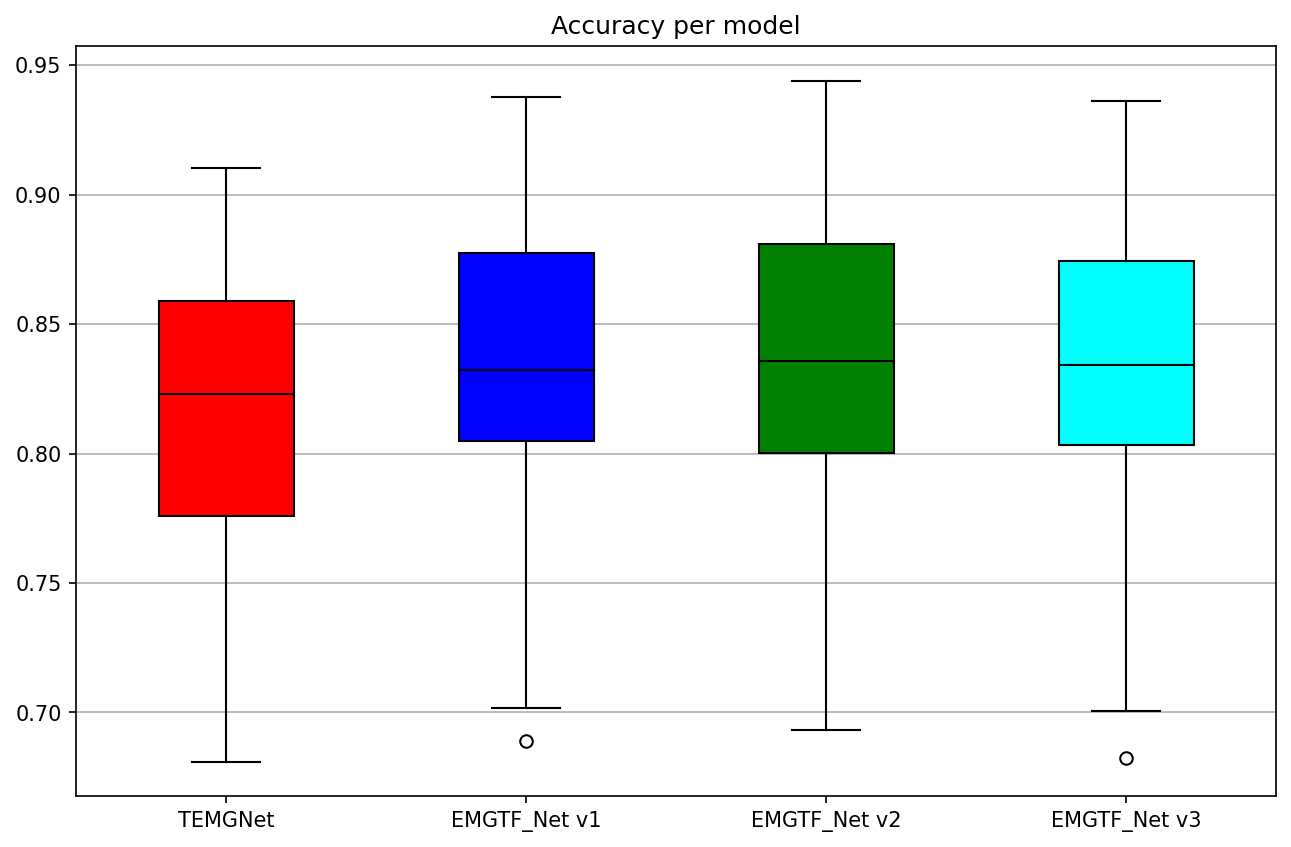}
    \caption{ Boxplot graph showing accuracy results per model }
    \label{fig:boxplot}
\end{figure}

\begin{figure*}
    \centering
    \includegraphics[width=0.75\linewidth]{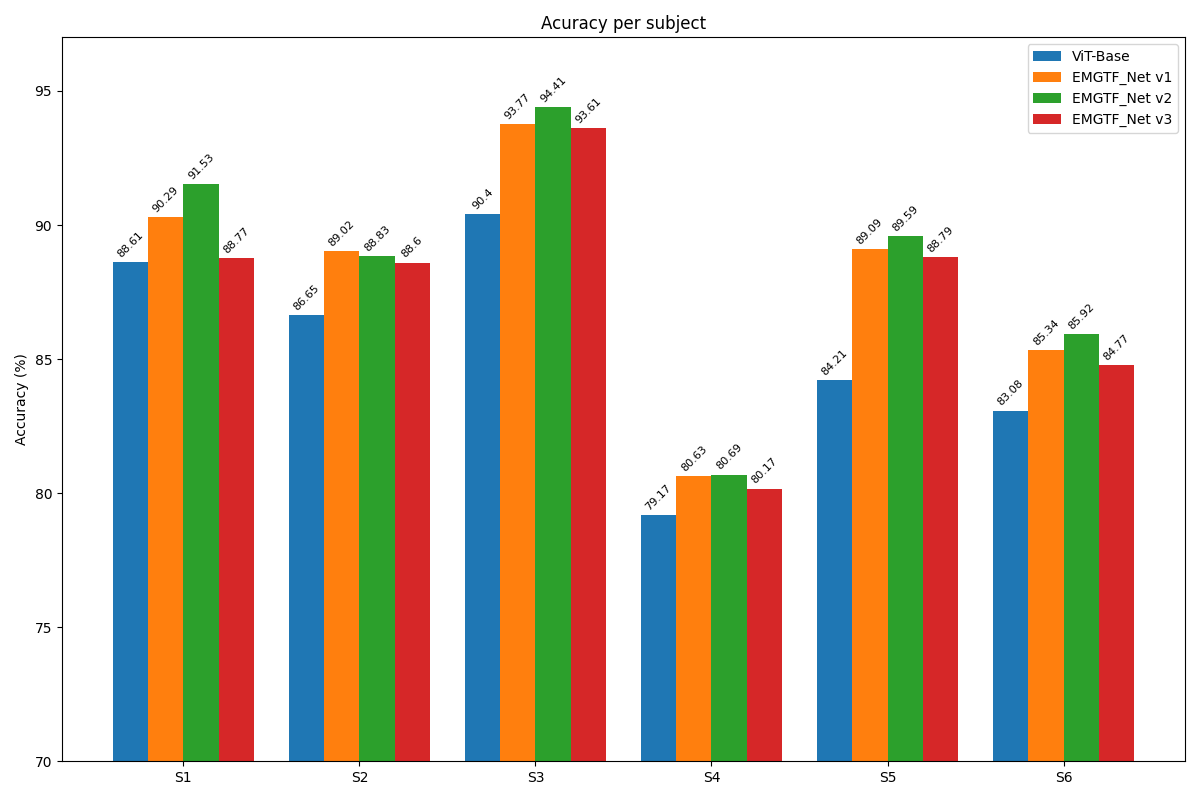}
    
    \caption{ Bar plot graph showing accuracy results per subject for each model in a subset of the DB2 }
    \label{fig:barplot}
\end{figure*}

The results of the classification accuracy performance are shown in a box plot in Fig. \ref{fig:boxplot}. The three versions of EMGTFNet outperformed the TEMGNet since their interquartile range (IQR) of classification accuracy are higher than the IQR of TEMGNet. Table 1 tabulates the classification accuracy of the three versions of EMGTFNet as well as our baseline model (TEMGNet) and other deep learning techniques of state-of-art.
%https://www.sciencedirect.com/science/article/pii/S0002939421003329?casa_token=IaaD8TIDT1gAAAAA:TROjkskehLuzTRLW1dyaPuGK0JBCJ2R4DxfJuiI_qjjpVpiuPb2RaL4svX-D8JVkCVmCiktuQw
%the network for segmenting OCT images. %is that the diference among Q3 and Q1. %Figure 2, B, shows the variation in Dice coefficients as a box plot with mean and standard deviation values as 0.90 ± 0.04.
% It demonstrates the efficacy of the network for segmenting OCT images.
The better result reported was reached by EMGTFNet v2
with 83,57\% classification accuracy. The bar plot in Fig. \ref{fig:barplot} shows the results per subject for each model in the first six subjects of the database. As we can see, for 5 of the 6 subjects chosen in the sample, EMGTFNet v2 reached a higher classification accuracy. In summary, the three versions EMGTFNet outperformed the classification accuracy from state-of-art. Hence, it demonstrates the efficacy of adding an FNB in the ViT architecture.

The findings of this study provide valuable insights into the application of Fuzzy Neural Block (FNB) with Vision Transformer (ViT) in one architecture called EMGTFNes for sEMG signal classification. Furthermore, the attained performance improvements demonstrate the potential of EMGTFNes in improving the accuracy of Hand Gesture Recognition. Likewise, including Neural Block (FNB) introduces fuzzy parameters to face the variability across time and uncertainty of sEMG signals. 

%\begin{table}[htbp]
%\caption{Table Type Styles}
%\begin{center}
%\begin{tabular}{|c|c|c|c|}
%\hline
%\textbf{Table}&\multicolumn{3}{|c|}{\textbf{Table Column Head}} \\
%\cline{2-4} 
%\textbf{Head} & \textbf{\textit{Table column subhead}}& \textbf{\textit{Subhead}}& \textbf{\textit{Subhead}} \\
%\hline
%copy& More table copy$^{\mathrm{a}}$& &  \\
%\hline
%\multicolumn{4}{l}{$^{\mathrm{a}}$Sample of a Table footnote.}
%\end{tabular}
%\label{tab1}
%\end{center}
%\end{table}

\begin{table}[h]
	%\captionsetup{font=footnotesize}
	%\footnotesize
	\centering
	\caption{A COMPARISON OF THE OVERALL ACCURACY OVER ALL PARTICIPANTS FOR EACH EMGTFNet MODEL.}
	\label{tab:table1}
	\begin{tabular}{cccc}

		\toprule

		Model & Window size & Average Accuracy & Parameters  \\ \midrule
		LSTM \cite{Sun_2021} & 300ms & 79.7 & 466,944  \\
		CNN \cite{Ding2018} & 200ms & 78.86 & -  \\
		LSTM-CNN \cite{Gulati2021} & 300ms & 81.96 & 1.1M  \\
		TEMGNet & 200ms & 81.67 & 54,609  \\
		EMGTF-Net v1  & 200ms & 83.48 & \textbf{56,793}  \\
		EMGTF-Net v2  & 200ms & \textbf{83.57} & \textbf{56,793}  \\
		EMGTF-Net v3  & 200ms & 83.30 & 57,561  \\

		\bottomrule
	\end{tabular}
\end{table}

\section{Conclusions}

In this work, we proposed a novel framework based on the addition of a Fuzzy Neural Block (FNB) in the transformer-based network for the task of Hand Gesture Recognition (HGR) from myoelectric signals. With the average accuracies achieved around 83.5\%, it can be concluded that the implementation of the FNB in the Transformer architecture increases the model's performance, overcoming state-of-the-art results in pattern recognition; thus demonstrating that FNB helps classify atypical and noisy values of the signal. Leveraging on the benefits of the Transformer, the proposed framework showed a higher accuracy with less than 60,000 parameters; accurately classifying a large number of hand gestures without any data augmentation technique. On the other hand, raw signals didn't prove to be a compatible input for the ViT architecture, so special signal processing had to be followed in order to achieve the desired results. The FNB showed promising results in deep learning applications for hand gesture recognition and should be further investigated in different and more complex architectures for future work. We intend to explore applications of this framework for bimanual task dexterity analysis \cite{kiani2019}, cognitive neuroscience \cite{andreu2021exp}, intelligent home interaction \cite{andreu2009, Vega_2022, cortez2020, kiani2022} and robotics \cite{andreu2018robots}.

\section*{Acknowledgment}
This work was partially supported by Fondo
Semilla, UTEC, (871059-2022). 
%\section*{References}

\bibliography{conference} 
\bibliographystyle{ieeetr}
\end{document}